# Target Attack Backdoor Malware Analysis and Attribution

Anthony Cheuk Tung Lai[1], Vitaly Kamluk[2], Alan Ho[1], Ping Fan Ke[3], Byron Wai[1]

**Abstract**

Backdoor Malware are installed by an attacker on the victim's server(s) for authorized access. A customized backdoor is weaponized to execute unauthorized system, database and application commands to access the user credentials and confidential digital assets. Recently, we discovered and analyzed a targeted persistent module backdoor in Web Server in an online business company that was undetectable by their deployed Anti-Virus software for a year. This led us to carry out research to detect this specific type of persistent module backdoor installed in Web servers. Other than typical Malware static analysis, we carry out analysis with binary similarity, strings, and command obfuscation over the backdoor, resulting in the Target Attack Backdoor Malware Analysis Matrix (TABMAX) for organizations to detect this sophisticated target attack backdoor instead of a general one which can be detected by Anti-Virus detectors. Our findings show that backdoor malware can be designed with different APIs, commands, strings, and query language on top of preferred libraries used by typical Malware.

## 1. Introduction

Backdoor Malware can exist on various devices ranging from a victim PC (which is compromised and become a bot controlled and accessible by the attacker), a server, an IoT device, or a mobile device. It can perform stealth activity including capturing the credentials, conversing with other systems and applications, and gathering more information to enable executing commands to query confidential databases and launching a further attack against other targets [1].

Recently, the threat of a backdoor developed with an IIS native module is increasing in prevalence [2]. We have been called for incident response and computer forensics by an online business company. We have found the IIS native module backdoor has been installed to the web server for at least six months without any detection until we have discovered the backdoor via the error logs, which is about failing to load the IIS native module backdoor with a compromised configuration file. The revealed backdoor comprises functions to access a database and modify table entries of a transaction. The backdoor allows an attacker to interact with it to execute various built-in SQL query commands against the application and database and database system administrator (SA) process take-over statement and create unauthenticated and authorized sensitive transaction records. In a typical Web application system architecture, we engage a three-tier design, and the data access layer is always deployed as a data access layer between a business logic layer in the Web application server and database [6]. It is not common to deploy at the Web Server layer (before the presentation layer). In certain scenarios, a native module is developed to have database tables or logs backup, temporary transaction table clean up and different database maintenance tasks. All existing controls including Windows Defender, Symantec Anti-Virus, and Firewall cannot detect the backdoor.

It is challenging to detect such stealth passive backdoor. Typical malware can be easily detected by anti-virus as most of the malware samples are circulated publicly. Malware analysts and anti-virus product vendors can extract signatures and bytecode patterns to update the virus definition file of the anti-virus software. Unfortunately, target attack backdoor malware is not widely circulated and it is customized and deployed as a business legitimate module to access data of an organization.

## 1. Background

[1] VX Research Limited. {darkfloyd, alanh0, byronwai}@vxrl.hk
[2] vitaly.kamluk@gmail.com
[3] Singapore Management University. pfke@smu.edu.sg

### Target Attack

Target Malware is called an Advanced Persistent Threat (APT), which targets a specific type of industry and/or an organization, aiming for espionage, financial interest, and service disruption. A Target Attack is well organized by a group of determined, well-focused, and coordinated attackers [3] and deliver the attack in various stages including reconnaissance (also known as information gathering), attack vector delivery, exploit and compromise the target and maintaining a persistent connection with the infected systems [4].

A backdoor can also be installed on a server at the system level or application level. In this paper, we will concentrate on Microsoft IIS (Internet Information Server) for our investigation. In the following sections, we discuss various types of backdoors, research motivation, and technical approaches to detect a backdoor.

### Backdoor Malware

A backdoor malware is an independent module or program that is not part of any installed modules. It is triggered as a backdoor when it receives the attacker's commands. In recent decades, attackers have always deployed backdoor with Webshell, which is a web application that allows them to interact with the victim server remotely. Researchers have designed a methodology [5] to detect them including identifying blacklisted keywords. However, it is treated as part of the web page under the "Web" page folder and the code can be obfuscated and encrypted. Meanwhile, some sophisticated Backdoors are developed with detailed system and operation knowledge of the victim organizations, even embedding specific commands and queries targeting the victim organizations' systems and applications. In this paper, we focus on this type of backdoor malware.

## 2. Challenges of Analysis and Detection

### Emerging Backdoor Threat

An attacker knows very well that if a backdoor to the web server is installed and it is impossible for the security administrator to discover because the Web server traffic is high. The context of request headers can be varied a lot, and it is challenging to identify malicious requests. Even in the best case, the security administrator can identify the attack requests, due to web server log limitations, it only captures simply GET requests in URI. As backdoor attacks are always communicating via HTTP POST request instead of GET request, it is impossible to have a full diagnosis of the malicious activity. Furthermore, to the best of our knowledge, recent studies of the IIS Native Module backdoor are not yet published. Moreover, there is a large target attack campaign, so-called "Advanced Persistent Threat (APT)" that has engaged the IIS native module as their backdoor to access victims' servers.

### Limitation of Dynamic Analysis

Most of the Malware can be analyzed via dynamic analysis, which means we execute the Malware and study the behavioral indicators; it is applicable to common Malware, unfortunately, for the specific dynamic loading module backdoor, we need a particular setup of a server to load them successfully. Once it runs, there will be no active footprints made by the backdoor as it waits for the command from the attacker. Malware analysts should reverse engineer the backdoor before carrying out any interaction with the backdoor, which depends on the complexity of the backdoor; reverse engineering can be very resource-intensive, which cannot immediately be helpful to incident response and detection.

### Limitation of Security Hardening, Audit, and Assessment

Researchers can argue that security or/and server administrators can identify the backdoor easily through regular audit of the installed module and set up a whitelist of legitimate modules. However, in our situation, the attacker can install and uninstall the malicious module backdoor through vulnerabilities or channel(s) unknown to us even latest system security vulnerabilities are already patched. In addition, the attacker can rename the modules as benign ones.

In view of the above challenges, we have considered these to develop further analysis to study these backdoor features, attribution, and family.

## 3. Methodology

We do not depend on public virus sandbox and scanning tools like Virus Total, because Malware targeting an organization is normally not circulated publicly.

In addition, most of the backdoor detection is dependent on network traffic analysis. We hope to feature the state-of-the-art of extracting more static features of Backdoor by referencing current and published backdoor malware and proof of concept backdoor, thereby developing a practical Target Attack Backdoor Malware Analysis Matrix (TABMAX), targeting to detect malicious and persistent module backdoor in both Apache and Microsoft platforms. The achievement done with this approach has identified several hidden backdoors in other networks of the same company with our analysis artifacts.

We have presented a TABMAX as below so as to effectively analyze this specific type of target attack backdoor malware:

| Task | Purpose | Description |
|---|---|---|
| **Static Analysis** | **Analysis and Attribution** | ● Analyze the malware without executing it.<br>● Understand the functions of malware.<br>● Reverse engineering into the Malware and find out any attribution information including URL(s), domain(s), and IP address(es). |
| **Binary Similarity** | **Analysis and Attribution** | ● Compare the malware samples with BinDiff and find out the difference between the malware variants.<br>● Compare the malware samples with SimHash and find out the basic function block similarities between malware. |
| **Strings and Obfuscation Detection** | **Detection** | ● Popular Command, Keywords, and API of Backdoor.<br>● Base64 encoded strings.<br>● Identification of HTTP Content-Type which is used to upload or download data.<br>● The number of SQL strings and obfuscation rate of SQL strings.<br>● The number of Powershell command strings and obfuscation rate of Powershell command strings. |
| **Assembly Instruction-based Detection** | **Detection** | The number of comparison instruction and extraction of the operands to get the command of Backdoor with given HTTP Header Content-Type(s). |

**Table: Target Attack Backdoor Malware Analysis Matrix (TABMAX)**

## 4. Malware Analysis

### Static Analysis

Under static analysis, we have investigated the backdoor Malware samples of OilRig (SHA256: 497e6965120a7ca6644da9b8291c65901e 78d302139d221fcf0a3ec6c5cf9de3) and uploaded the samples in two different time slots to check the detection status among anti-virus detectors and YARA rules (Fig. 3a and Fig. 3b). It is found more detectors and YARA rules can find out the OilRig backdoor.

Meanwhile, we have done the same approach over two backdoor malware from our online business customer. There is no anti-virus detector and YARA rules that find out the target attack backdoor (Fig. 3c and Fig. 3d). One of the versions of the backdoor malware exhibits its original filename is *acproxy*, which matches our first discovery at the server (Fig. 3e).

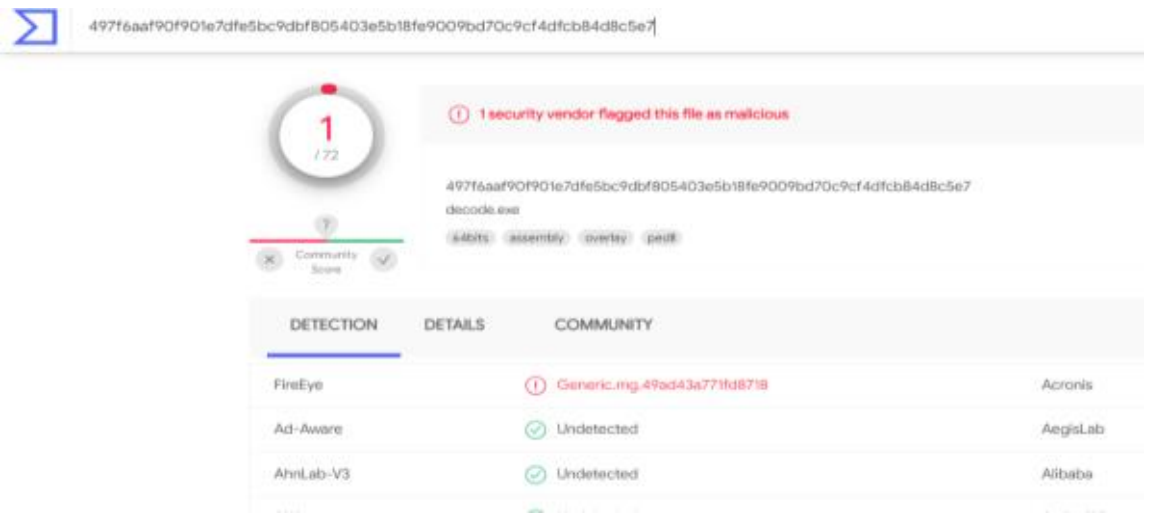

**Fig. 3a.** We have uploaded another published backdoor DLL file (SHA-256: 497e6965120a7ca6644da9b8291c65901e78d302139 d221fcf0a3ec6c5cf9de3) from OilRig APT campaign to VirusTotal and there is one out of seventy-two anti-virus software can detect it on 30 June 2020.

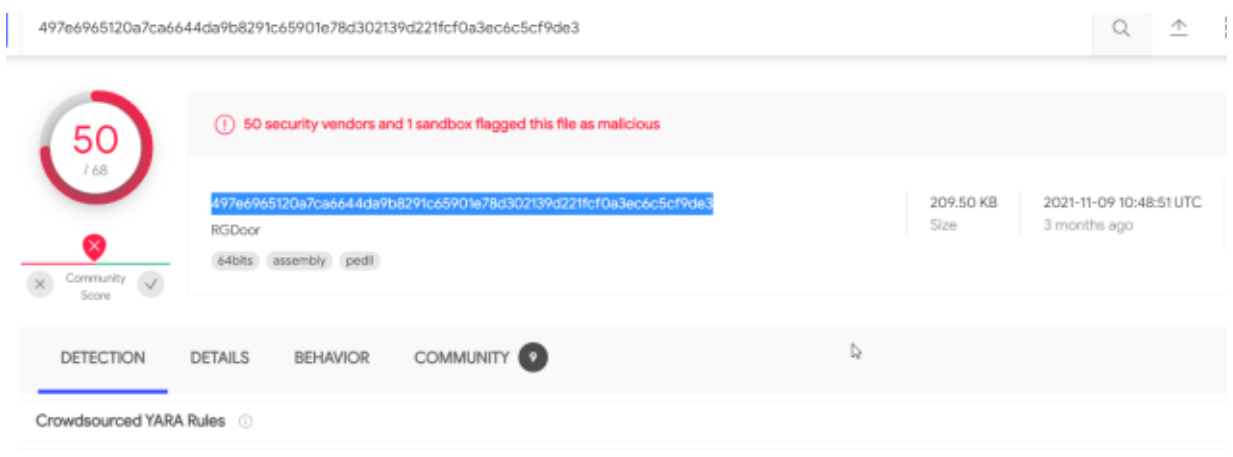

**Fig. 3b.** We have uploaded another published backdoor DLL file (SHA-256: 497e6965120a7ca6644da9b8291c65901e78d302139 d221fcf0a3ec6c5cf9de3) from OilRig APT campaign to VirusTotal and there is fifty out of sixty eight anti-virus software can detect it on 4 February 2022.

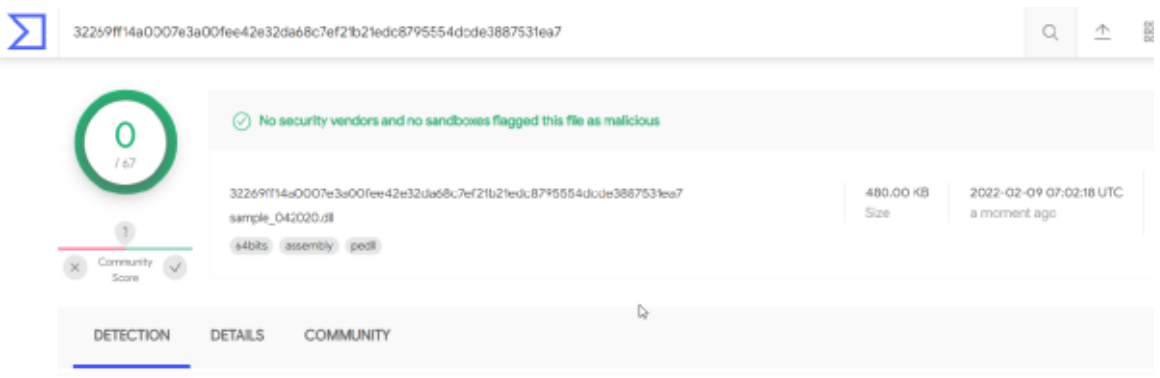

**Fig. 3c.** We have uploaded the backdoor DLL file with file name Transtatic.dll found in April 2020 (SHA-256: 32269ff14a0007e3a00fee42e32da68c7ef21b21edc8795554dcde3887531ea7) to VirusTotal and none of the anti-virus software and Yara rules can detect it.

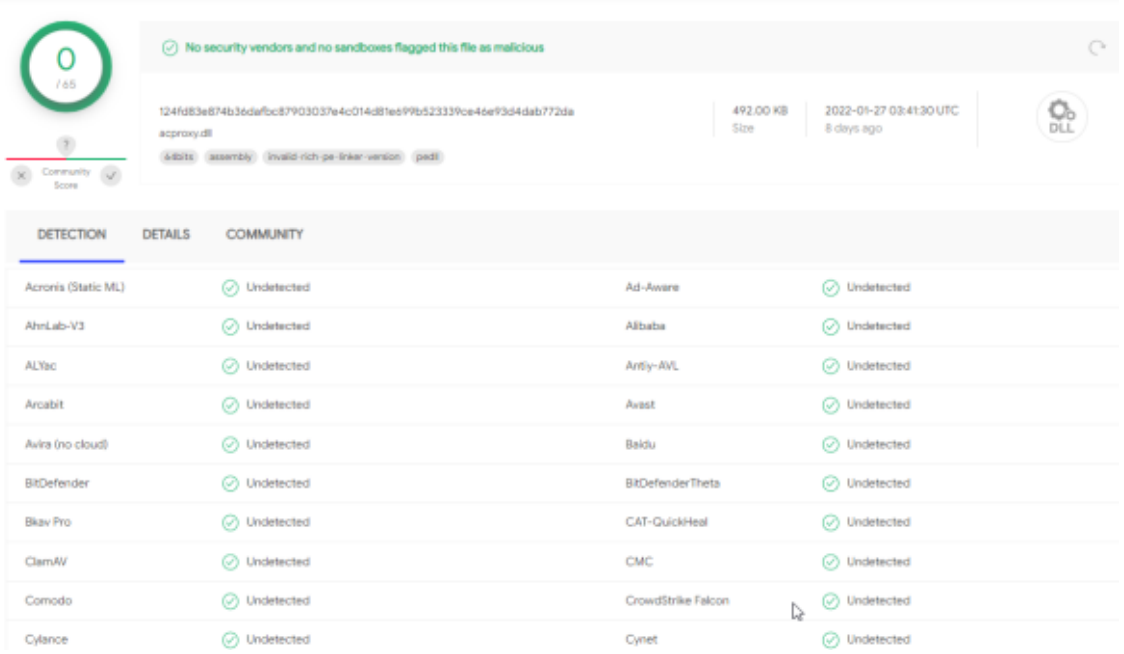

**Fig. 3d.** We have uploaded the backdoor DLL file with file name Transtatic.dll found in September 2020 (SHA-256: 124fd83e874b36dafbc87903037e4c014d81e699b523339ce46e93d4dab772da) to VirusTotal and none of the anti-virus software and Yara rules can detect it.

**File Version Information**

| | |
|---|---|
| Copyright | Copyright (C) 2008 |
| Product | Microsoft Windows Operating System |
| Description | Autochk Proxy DLL |
| Original Name | acproxy.dll |
| Internal Name | acproxy.dll |
| File Version | 10.0.18362.1 |

**Fig. 3e.** The original name of the backdoor malware filename is acproxy.dll.

We attempt to check the function capability of our backdoor malware samples with capa[]. Several encoding and hashing schemes are identified but those listed capabilities and details cannot conclude the malware is a backdoor (Fig. 3f).

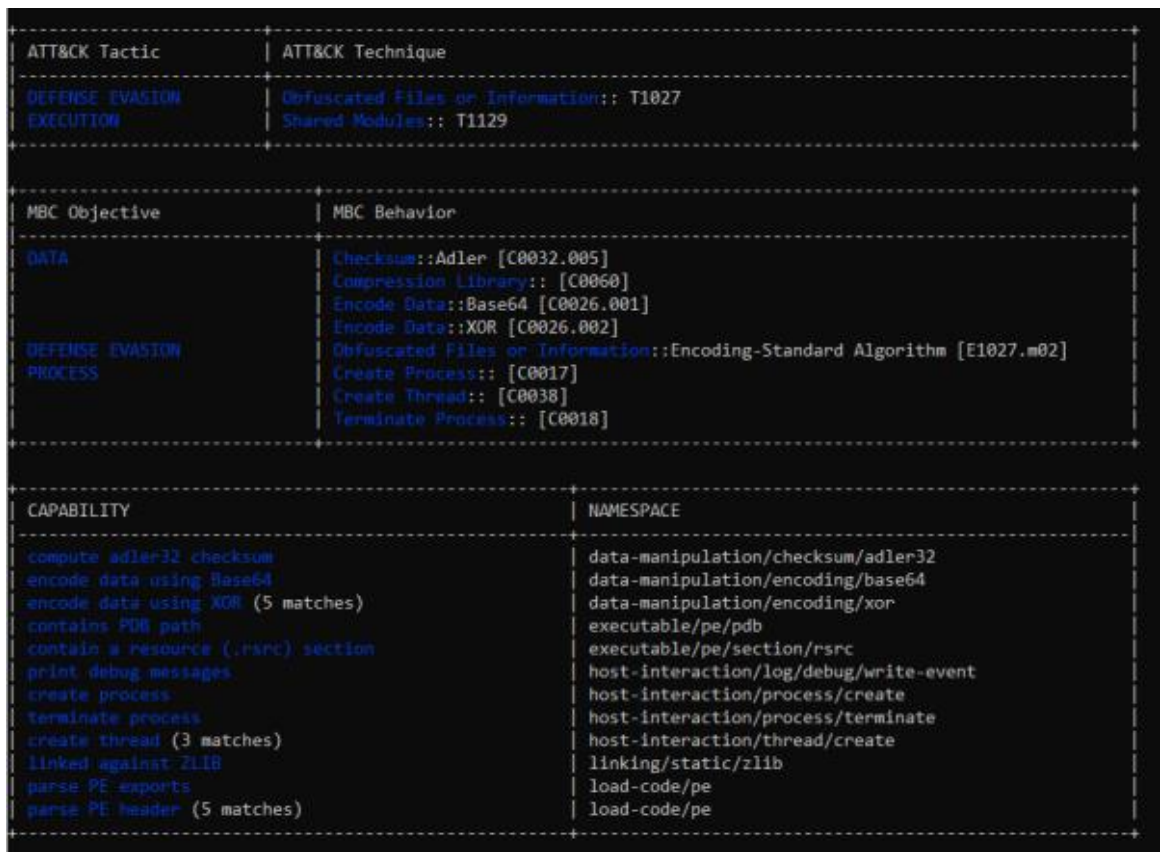

| ATT&CK Tactic | ATT&CK Technique |
|---|---|
| DEFENSE EVASION | Obfuscated Files or Information:: T1027 |
| EXECUTION | Shared Modules:: T1129 |

| MBC Objective | MBC Behavior |
|---|---|
| DATA | Checksum::Adler [C0032.005] |
| | Compression Library:: [C0060] |
| | Encode Data::Base64 [C0026.001] |
| | Encode Data::XOR [C0026.002] |
| DEFENSE EVASION | Obfuscated Files or Information::Encoding-Standard Algorithm [E1027.m02] |
| PROCESS | Create Process:: [C0017] |
| | Create Thread:: [C0038] |
| | Terminate Process:: [C0018] |

| CAPABILITY | NAMESPACE |
|---|---|
| compute adler32 checksum | data-manipulation/checksum/adler32 |
| encode data using Base64 | data-manipulation/encoding/base64 |
| encode data using XOR (5 matches) | data-manipulation/encoding/xor |
| contains PDB path | executable/pe/pdb |
| contain a resource (.rsrc) section | executable/pe/section/rsrc |
| print debug messages | host-interaction/log/debug/write-event |
| create process | host-interaction/process/create |
| terminate process | host-interaction/process/terminate |
| create thread (3 matches) | host-interaction/thread/create |
| linked against ZLIB | linking/static/zlib |
| parse PE exports | load-code/pe |
| parse PE header (5 matches) | load-code/pe |

**Fig. 3f. Function Capability Analysis of Backdoor**

Afterward, we carry out in-depth reverse engineering and summarise the following similarity between OilRig and Transtatic:

1. They support upload, download, and command execution functions. OilRig exhibits upload, download, and cmd as keywords to upload (Figure 4a). The Backdoor writer of Transtatic.dll manipulates a stealthier approach by taking Content-Type in HTTP header [7] with “IMAGE/type” as png, jpg, and gif so as to upload data to, download data from, and execute a command against the victim IIS Web server (Figure 4b).

2. Response data is sent with the Content-Type “Text/Plain” (Figure 4b)

3. When both Malware uploads the data, they will detect the end of upload completion if the error or failure message is found after a while loop of data transfer (Figure 5a and Figure 5b). Transtatic backdoor will use GetTickCount function to count the time when uploading the data, which is also the Malware writer’s preference in calculating the time (Figure 5c).

4. 4. Keywords and API used in both Backdoor samples include *RegisterModule*, *CHttpModule*, *“Server Error”*, *“Failed”*, etc.

```
  if ( v89 < 0x10 )
    goto LABEL_104;
  v32 = Memory;
  goto LABEL_103;
}
if ( strstr((const char *)v15, "upload$") )
{
  v33 = (const char *)(*(__int64 (__fastcall **)(__int6
                       v5,   |
                       "Content-Length",
                       &v83);
  v34 = v33;
  if ( v83 && atoi(v33) )
```

**Figure 4a: Upload data command in the OilRig backdoor.**

```
v9 = *(_BYTE **)(v7 + 8);
if ( *v9 != 'I' || v9[1] != 'M' || v9[2] != 'A' || v9[3] != 'G' || v9[4] != 'E' || v9[5] != '/'
  return 0i64;
v10 = v9[6];
if ( v10 == 'j' )
{
  if ( v9[7] == 'p' && v9[8] == 'g' )
  {
    sub_180002360(v9, (unsigned int)v8, &v34);
```

**Figure 4b: Stealthy approach of data upload, data download, and command execution in Transtatic.dll.**

```
v77 -= 0xFFFF;
if ( v78 < 0 )
{
  LODWORD(v80) = v78;
  (*(void (__fastcall **)(__int64, signed __int64,
    v76,
    500i64,
    "Server Error",
    0i64,
    v80,
    0i64,
    0i64);
}
```

**Figure 5a: Provide "Server Error" message when data is downloaded in OilRig backdoor.**

```
  ++v14;
while ( *((_BYTE *)&v34 + v14) );
v30 = (unsigned __int16)v14;
v15 = (*(__int64 (__fastcall **)(__int64, int *, si
if ( v15 < 0 )
  (*(void (__fastcall **)(__int64, signed __int64,
    v3,
    500i64,
    "Server Error",
    0i64,
    v15,
    0i64,
    0i64);
return 2i64;
```

**Figure 5b: Provide "Server Error" message when data is downloaded in Transactic.dll.**

```
if ( v10 != 'p' || v9[7] != 'n' || v9[8] != 'g' )
  return 0i64;
v17 = 0i64;
v18 = sub_180039BD0(v9 + 9);
EnterCriticalSection(&stru_180078438);
v19 = qword_180078B48;
v20 = xmmword_180078B50;
v21 = (_DWORD **)qword_180078B48;
if ( qword_180078B48 != (_QWORD)xmmword_180078B50 )
{
  while ( 1 )
  {
    v22 = (signed __int64)(v21 + 1);
    if ( **v21 == v18 )
      break;
    ++v21;
    if ( v22 == (_QWORD)xmmword_180078B50 )
      goto LABEL_47;
  }
  v17 = *v21;
  sub_180034590(v21, v21 + 1, xmmword_180078B50 - v22);
  v19 = qword_180078B48;
  v20 = xmmword_180078B50 - 8;
  *(_QWORD *)&xmmword_180078B50 = xmmword_180078B50 - 8;
LABEL_47:
  if ( v19 != v20 )
  {
    v23 = v19 + 8;
    do
    {
      if ( GetTickCount() - *(_DWORD *)(*(_QWORD *)v19 + 80i64) < 0xEA60 )
```

**Figure 5c. Use of GetTickCount in the upload function. The attacker enters IMAGE/png as the command to upload the data to the backdoored IIS Web server.**

On the other hand, we attempt to find the following similarity between two variants of our Transtatic samples. We have identified the malware has changed the tempdb name from *salsync* to *dynupper* in their every SQL queries which attempts to modify the transactions in the database (Fig. 5d). We have tried to search from Internet and different virus repositories but there is no connection to any current published malware attribution. At the same time, we have identified company name like dyn from a social media portal about job review (Fig. 5e).

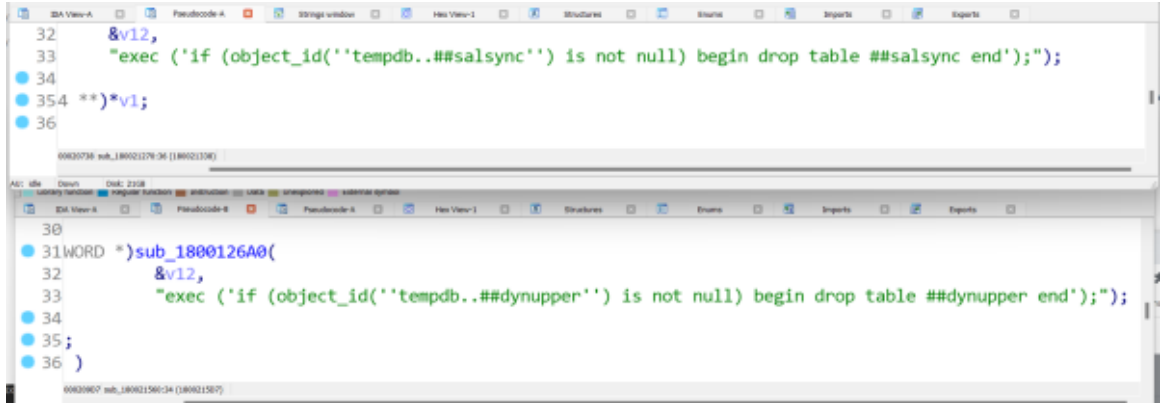

**Figure 5d. The tempdb name is changed from salsync to dynupper between two Transtatic backdoor samples in April and September 2020 respectively.**

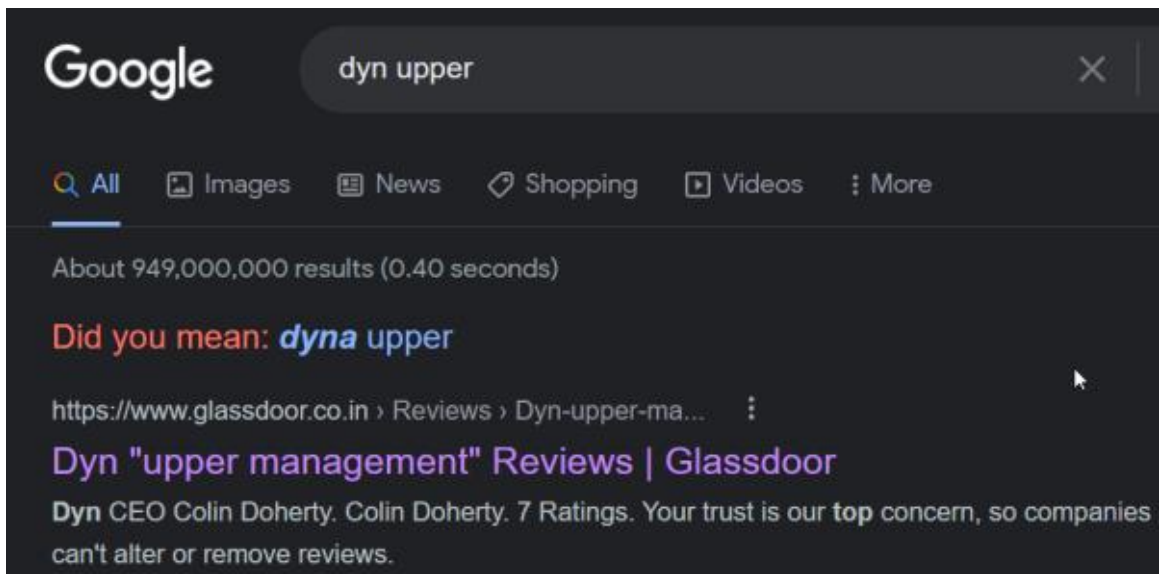

**Figure 5e. The tempdb name is changed from salsync to dynupper between two Transtatic backdoor samples in April and September 2020 respectively.**

## Binary Similarity

As this backdoor is a targeted attack, we attempt to find any variant, similar or published backdoor for our comparative study if we need to extend our attribution analysis. In the following section, we can successfully identify several difference between two different version our backdoor malware variants and similarities of our backdoor sample with another APT group FunnyDream.

To begin with, we use BinDiff tool[8] to identify 58 unmatched functions (Fig. 6a) of the variants and manually review those functions. In our finding, we have identified a new database data and record process function (Fig. 6b) and network data streaming and buffer handling function (Fig. 6c).

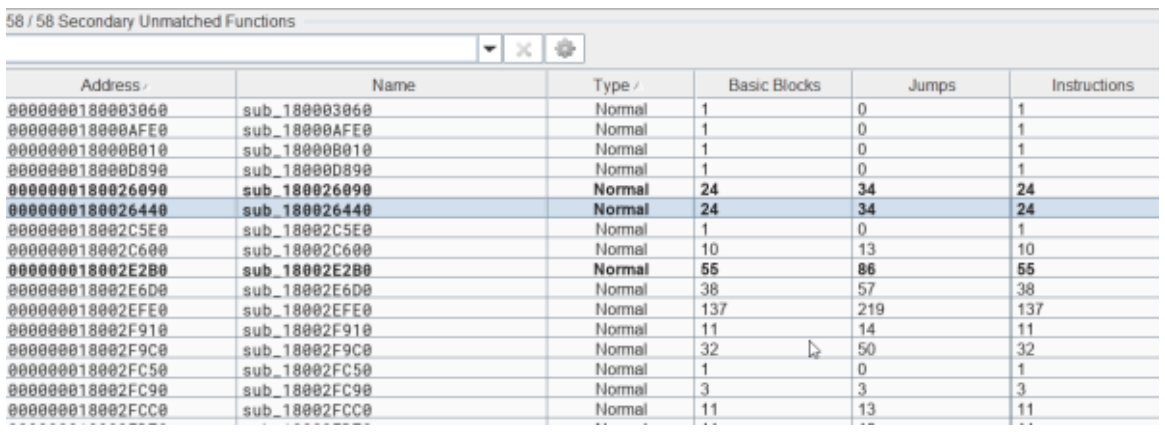

58 / 58 Secondary Unmatched Functions

| Address | Name | Type | Basic Blocks | Jumps | Instructions |
|---|---|---|---|---|---|
| 0000000180003060 | sub_180003060 | Normal | 1 | 0 | 1 |
| 000000018000AFE0 | sub_18000AFE0 | Normal | 1 | 0 | 1 |
| 000000018000B010 | sub_18000B010 | Normal | 1 | 0 | 1 |
| 000000018000D890 | sub_18000D890 | Normal | 1 | 0 | 1 |
| 0000000180026090 | sub_180026090 | Normal | 24 | 34 | 24 |
| 0000000180026440 | sub_180026440 | Normal | 24 | 34 | 24 |
| 000000018002C5E0 | sub_18002C5E0 | Normal | 1 | 0 | 1 |
| 000000018002C600 | sub_18002C600 | Normal | 10 | 13 | 10 |
| 000000018002E2B0 | sub_18002E2B0 | Normal | 55 | 86 | 55 |
| 000000018002E6D0 | sub_18002E6D0 | Normal | 38 | 57 | 38 |
| 000000018002EFE0 | sub_18002EFE0 | Normal | 137 | 219 | 137 |
| 000000018002F910 | sub_18002F910 | Normal | 11 | 14 | 11 |
| 000000018002F9C0 | sub_18002F9C0 | Normal | 32 | 50 | 32 |
| 000000018002FC50 | sub_18002FC50 | Normal | 1 | 0 | 1 |
| 000000018002FC90 | sub_18002FC90 | Normal | 3 | 3 | 3 |
| 000000018002FCC0 | sub_18002FCC0 | Normal | 11 | 13 | 11 |

**Figure 6a. Identify unmatched functions between our two backdoor transtatic samples**

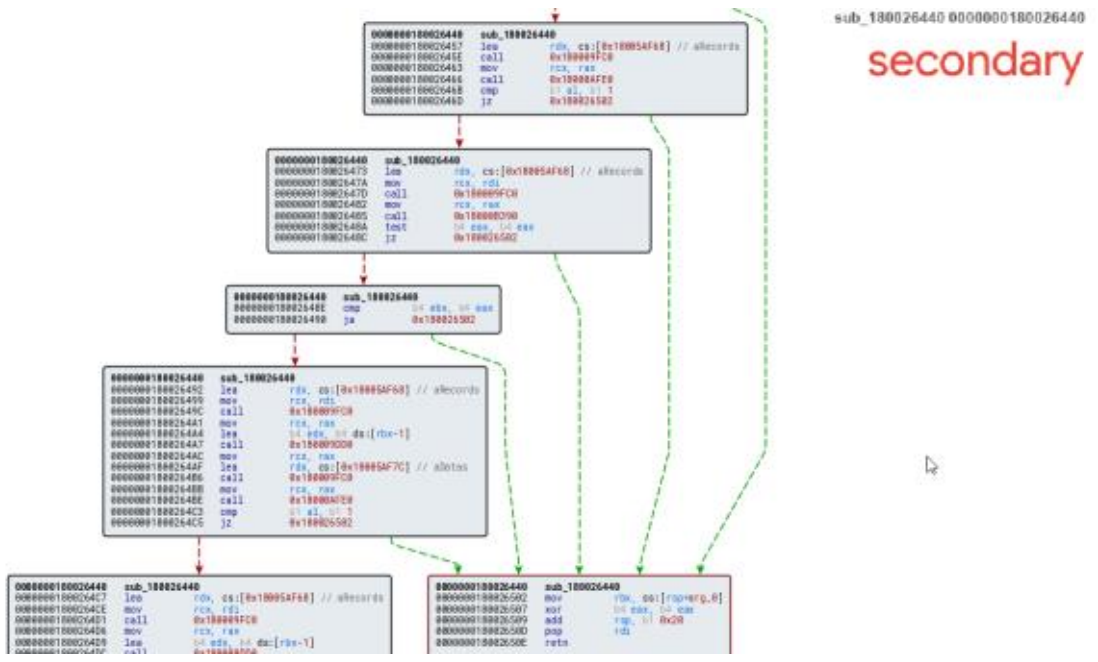

**Figure 6b. A new data and record processing function at sub_180026440 found in the recent transtatic variant. (SHA-256: 124fd83e874b36dafbc87903037e4c014d81e699b52 3339ce46e93d4dab772da)**

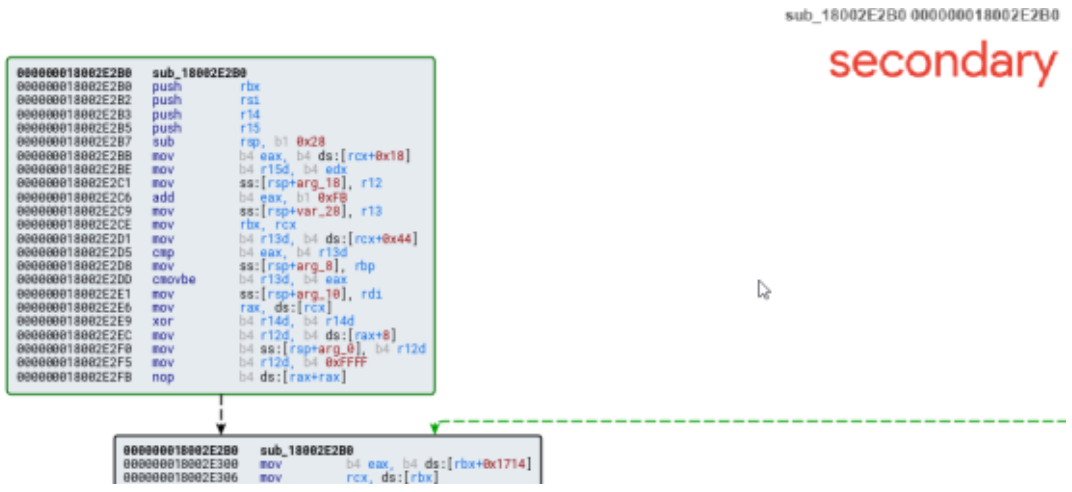

**Figure 6c. A data stream transmission and buffer function at sub_18002e2b0 found in the recent transtatic variant.**

As this backdoor is a targeted attack, we attempt to find any similar or published backdoor for our comparative study if we need to undergo further forensic and investigation. In the following section, we can successfully identify several similarities of our backdoor sample with another APT groups called FunnyDream. From our similarity check, we can identify fully matched functions, however, those functions are mostly library functions. In the following table with excerpts, we can only identify a function sub_180030760 which is a malware user code. The function is about alder32 hashing function. This hash function is identified with capa, however, we have found the highlighted value 347 is exclusively found Alder32 function in FunnyDream family with verification with existing malware sample pool with Virus Hunt.

```
[!] (347/760 - 13 branching nodes)
0.976562: 124fd83e874b36da.180030760
matches 28570122e952f25c.18000c4d0
[!] (347/760 - 13 branching nodes)
0.976562: 124fd83e874b36da.180030760
matches c2dbaafccfb8c912.180001070
[!] (347/760 - 13 branching nodes)
0.976562: 124fd83e874b36da.180030760
matches 9d0c2e8d0e2430c3.14000c5e0
[!] (347/760 - 13 branching nodes)
0.976562: 124fd83e874b36da.180030760
matches b3c811595a0edbed.180001070
[!] (347/760 - 13 branching nodes)
0.976562: 124fd83e874b36da.180030760
matches 6543180ba4e195b4.140001070
[!] (347/760 - 13 branching nodes)
0.976562: 124fd83e874b36da.180030760
matches 8f79333f2cc38d22.14000c4d0
[!] (347/760 - 13 branching nodes)
0.875000: 124fd83e874b36da.180030760
```

```
matches 2e6dfca6b2b8a11d.40ae10
[!] (347/760 - 13 branching nodes)
0.875000: 124fd83e874b36da.180030760
matches 1aa170383d473b6a.4062d0
[!] (347/760 - 13 branching nodes)
0.875000: 124fd83e874b36da.180030760
matches 631c62e067667a02.1000b1c0
[!] (347/760 - 13 branching nodes)
0.867188: 124fd83e874b36da.180030760
matches a1859ce1575ab08b.10001060
[....]
```

Table: Simhash similarity between Transtatic sample and FunnyDream samples

```
__int64 __fastcall sub_180030760(unsigned int a1, unsigned __int8 *a2, unsigned __int64 a3)
{
 [....]
 a3 = (unsigned int)a3;
 v3 = a2;
 v4 = HIWORD(a1);
 a1 = (unsigned __int16)a1;
 if ( (unsigned int)a3 == 1i64 )
 {
  v5 = (unsigned __int16)a1 + *a2;
  v6 = v5 - 65521;
  if ( v5 < 0xFFF1 )
   v6 = v5;
  v7 = v6 + v4 - 65521;
  if ( v6 + v4 < 0xFFF1 )
   v7 = v6 + v4;
  return v6 | (v7 << 16);
 }
 else if ( a2 )
 {
  if ( (unsigned int)a3 >= 0x10ui64 )
  {
   if ( (unsigned int)a3 >= 0x15B0ui64 )
   {
    v11 = (unsigned int)a3 / 0x15B0ui64;
    a3 = (unsigned int)a3 % 0x15B0ui64;
    do
    {
     v12 = 347;
     do
     {
      v13 = *v3 + a1;
      v14 = v13 + v4;
      [....]
      v73 = v3[14] + v71;
      v74 = v3[15];
      v75 = v73 + v72;
      v3 += 16;
      a1 = v74 + v73;
      v4 = a1 + v75;
      --v44;
     }
     while ( v44 );
    }
    for ( ; a3; --a3 )
    {
     v76 = *v3++;
     a1 += v76;
     v4 += a1;
    }
    a1 %= 0xFFF1u;
    v4 %= 0xFFF1u;
   }
   return a1 | (v4 << 16);
  }
  else
  {
   if ( (_DWORD)a3 )
   {
    do
    {
     v9 = *v3++;
     a1 += v9;
     v4 += a1;
     --a3;
    }
    while ( a3 );
   }
   v10 = a1 - 65521;
   if ( a1 < 0xFFF1 )
    v10 = a1;
   return v10 | ((v4 + 15 * (v4 / 0xFFF1)) << 16);
  }
 }
 else
 {
  return 1i64;
 }
}
```

**Figure 6d. Customized Alder32 function. [Need to see whether we take it off or not]**

## 5. Detection Methodology

Given the characteristics of the backdoor malware, we have provided TABMAX such that security administrators and officers can reference it. We have two approaches to giving the indicators in terms of Signature-based and Feature-based detection. We have successfully assisted another enterprise to detect insider deployed backdoor in both Microsoft and Apache platforms with TABMAX.

In TABMAX, we developed a scanner tool to identify any potential backdoor with the following functions.

**String and Obfuscation Detection**

Due to the assumption and limitation of our scanner, we do not attempt to give a clear-cut indicator or detection classifying whether it is a backdoor or not. We have provided different indicators to the security professionals to decide whether they should carry out further analysis of a particular DLLs via checking with an online virus checking sandbox (e.g., virus total) or carry out manual reverse engineering and malware analysis. We require one to review both numbers and the context of results. The idea is more indicators the native module matches, the higher probability the native module is malicious.

## I. Features

The scanner has the following functions for string-based obfuscation detection and Assembly instruction-based detection based on our investigation and correlation of our findings:

Other than the keyword and string pattern matching, we have attempted to identify whether the SQL strings and Powershell scripts are obfuscated through cosine similarity comparison between vectors of collected strings from SQL/Powershell script repositories and the target native module DLL file. The Cosine similarity is used in detecting obfuscated code in viruses [9] so as to detect virus variants more effectively. The more obfuscated strings of SQL and Powershell scripts are found, the higher the probability it is a backdoor Malware. Here is our methodology:

1. Build the frequency table of characters over files. By using the tool, we generate two standard Top 30 Character Frequency tables (Figure 7a and Figure 7b) and attached full tables in Appendix, one for SQL files and the other one for PowerShell files. The legitimate SQL and Powershell scripts files are collected from Microsoft's [10] and Azure's [11] Github repositories, respectively.
2. Extract the strings from a target file by using "Strings". It will scan a file for Unicode or Ascii strings of a default length of 3 or more characters.
3. Classify the strings by finding the language keywords. If the string contains SQL language keywords, it will be classified as SQL language if it also contains PowerShell language.
4. Build the frequency table of the classified strings.
5. Calculate the cosine similarity of the frequency table of the classified strings and the frequency table of the standard language files.
6. Find all base64 encoded strings.
7. Find the specific keywords, such as *"DownloadString", "upload", "download", "chttpmodule", "encode", "decode".*

| Character | Frequency |
|---|---|
| | |
| 0 | 0.087421 |
| <Space > | 0.075127 |
| e | 0.047623 |
| a | 0.034788 |
| 3 | 0.034726 |
| 6 | 0.034533 |
| C | 0.033376 |
| 4 | 0.033155 |
| , | 0.032057 |
| 9 | 0.032014 |
| t | 0.031492 |
| 1 | 0.025061 |
| i | 0.024927 |
| r | 0.022208 |
| o | 0.021458 |
| n | 0.021324 |
| 5 | 0.019975 |
| ' | 0.019325 |
| c | 0.019143 |
| d | 0.017487 |
| D | 0.017420 |
| 7 | 0.016492 |
| 2 | 0.016190 |
| F | 0.015992 |
| E | 0.013884 |
| 8 | 0.013748 |
| m | 0.012664 |
| l | 0.010761 |
| T | 0.010660 |
| S | 0.010444 |

| Character | Frequency |
|---|---|
| <Space> | 0.168742 |
| e | 0.085947 |
| t | 0.056438 |
| r | 0.055300 |
| o | 0.052609 |
| a | 0.048058 |
| i | 0.038527 |
| s | 0.036009 |
| n | 0.035964 |
| c | 0.030310 |
| - | 0.025493 |
| u | 0.025067 |
| m | 0.020010 |
| p | 0.019976 |
| l | 0.019949 |
| $ | 0.018316 |
| d | 0.016633 |
| g | 0.013452 |
| " | 0.012247 |
| N | 0.011838 |
| S | 0.010192 |
| A | 0.010040 |
| h | 0.009941 |
| . | 0.008611 |
| y | 0.007967 |
| b | 0.007843 |
| f | 0.007659 |
| I | 0.007238 |
| P | 0.006948 |
| R | 0.006778 |

**Figure 7b. Top-30 Character Frequency table for Powershell files.**

| **Figure 7a. Top-30 Character Frequency table for SQL file.** | |
|---|---|

## II. Experiments

We have scanned our current backdoor samples and legitimate IIS native module DLL files under the IIS system folder.

From Table I, we can find both backdoor malware hit a high number of keyword and API indicators, SQL strings, and Powershell script strings. More importantly, the obfuscation index on both SQL and Powershell scripts exhibits significant deviation between malware backdoor and the legitimate native module. However, standard Microsoft native modules, CustomError and HTTPCache, have been found indicators are similar to backdoor behaves. This requires further check over binary signature and hash for verification whether an existing library is backdoored or not.

| **Native Module Filename** | **No. of SQL strings** | **No. of ps1 strings** | **No. of Interesting strings /keywords and API** | **No. of Base64 encoded strings** | **Obfuscation index SQL** | **Obfuscation index ps1** |
|---|---|---|---|---|---|---|
| Transtatic.dll (Backdoor) (SHA256 Cannot be disclosed) | 78 | 666 | 5 | 79 | 0.889 | 0.983 |
| OilRig (Backdoor) (SHA256 497e6965120a7ca6644da9b8291c65901e78d302139d221fcf0a3ec6c5cf9de3) | 82 | 161 | 32 | 12 | 0.787 | 0.938 |
| FunnyDream (Backdoor) (SHA256: ) | 68 | 621 | 5 | 63 | 0.869 | 0.979 |
| IIS RAID(Backdoor) (SHA256 497e6965120a7ca6644da9b8291c65901e78d302139d221fcf0a3ec6c5cf9de3) | 13 | 109 | 2 | 3 | 0.652 | 0.834 |
| AnonymousAuthenticationModule | 5 | 25 | 8 | 0 | 0.649 | 0.731 |
| CustomError Module | 2 | 52 | 3 | 4 | 0.625 | 0.597 |
| DefaultDocument Module | 5 | 9 | 3 | 1 | 0.625 | 0.789 |
| DirectoryListing Module | 4 | 13 | 3 | 2 | 0.620 | 0.730 |
| HTTPCacheModule | 5 | 48 | 3 | 2 | 0.627 | 0.599 |
| HTTPLoggingModule | 4 | 34 | 3 | 2 | 0.645 | 0.805 |
| Protocol Support Module | 4 | 11 | 3 | 1 | 0.624 | 0.804 |
| Requeste Filtering Module | 21 | 26 | 4 | 0 | 0.650 | 0.863 |
| StaticCompressionModule | 2 | 26 | 3 | 5 | 0.624 | 0.730 |
| StaticFileModule | 8 | 29 | 3 | 4 | 0.628 | 0.749 |

**Table I: String and Obfuscation Detection.**

We have tried to upload the IIS RAID backdoor, which is an open-source backdoor in Github, to an online virus sandbox called VirusTotal. From the below figure, even IIS RAID is with obvious indicators revealed by our scanner by identifying string comparison functions and retrieving the corresponding argument with DMP and CMD commands, scanning result in VirusTotal [12] cannot show this kind of relevant information to the target organization for reference, and most of the leading anti-virus vendor engines cannot detect it. Others will argue whether we can simply whitelist the legitimate module to detect the malicious one. However, our situation and assumption are that the environment has no baseline of legitimate module whitelist is made and the attacker has compromised the system and he/she can rename the module looks like a legitimate one.

```
0x1800146ddL call    cs:CompareStringA
0x1800146ddL lea     rax, String2; "CMD|"
0x180014728L call    cs:CompareStringA
0x180014728L lea     rax, aPin; "PIN|"
0x180014779L call    cs:CompareStringA
0x180014779L lea     rax, aInj; "INJ|"
0x1800147c1L call    cs:CompareStringA
0x1800147c1L lea     rax, aDmp; "DMP|"
Total number of CompareStringA instructions: 4
```

**Figure 8c. Detect commands of IIS RAID backdoor.**

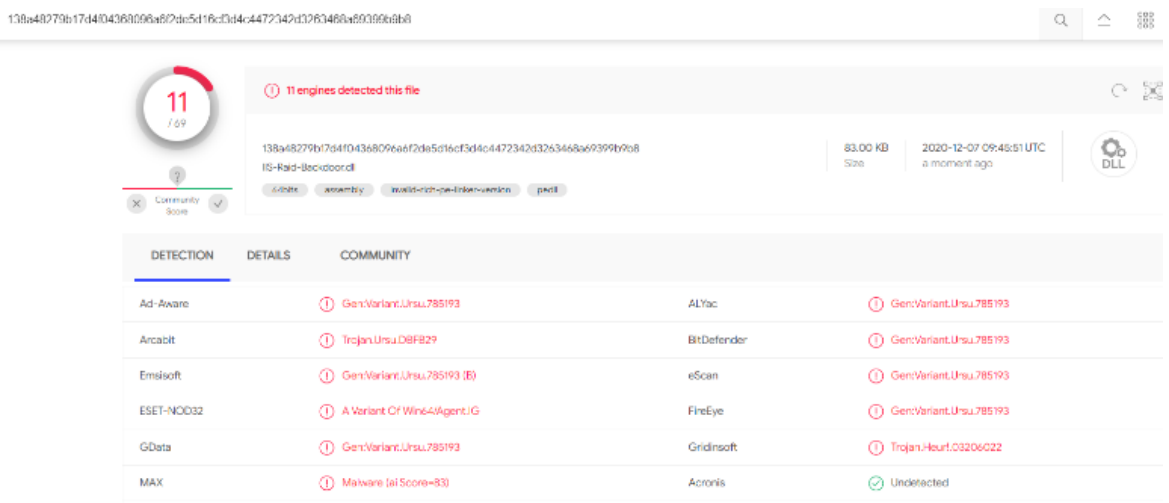

**Figure 8d. VirusTotal Scan of IIS RAID backdoor.**

### Assembly Instruction-based Detection

Previously, we have observed the outbound data is exported via "text/plain" with command strings only. In our scanner, we suggested the following instruction-based detection methodology to provide a backdoor indicator, as the backdoor needs a command to trigger every action and upload and download the target victim information at the Web server:

1. Search "text/plain" and other content-type values in the target IIS native module DLL file.

2. Locate the function that contains the reference to the "text/plain" or other content-type values string.

3. Identify all *cmp* instructions and *strstr* instructions that are used for comparison with operands with ASCII character code only. For Windows binaries, we will identify invocations of the *CompareStringA* function.

4. If high numbers of *cmp*, *strstr*, and *CompareStringA* are found under a data export function, we proposed this is an indicator of a backdoor from our surveying into benign and malicious modules.

| Native Module Filename | No. of CMP | No. of StrStr | No. of Calling Compare StringA | Identified Character Sequence and Command |
|---|---|---|---|---|
| Transtatic.dll (Backdoor) | 25 | 0 | 0 | Yes |
| OilRig (Backdoor) | 32 | 4 | 0 | Yes |
| FunnyDream (Backdoor) | 37 | 0 | 0 | Yes |
| IIS RAID(Backdoor) | 0 | 0 | 4 | Yes |
| AnonymousAuthenticationModule | 0 | 0 | 0 | No |
| CustomErrorModule | 0 | 0 | 0 | No |
| DefaultDocumentModule | 0 | 0 | 0 | No |
| DirectoryListingModule | 0 | 0 | 0 | No |
| HTTPCacheModule | 0 | 0 | 0 | No |
| HTTPLoggingModule | 0 | 0 | 0 | No |
| ProtocolSupportModule | 0 | 0 | 0 | No |
| RequesteFilteringModule | 0 | 0 | 0 | No |
| StaticCompressionModule | 0 | 0 | 0 | No |
| StaticFileModule | 0 | 0 | 0 | No |

**Table II: Assembly Instruction-based Detection.**

We have included a summary table of indicators (Table II) for both malicious and legitimate native modules. We have found that numbers for malicious backdoor with higher numbers in *CMP*, *StrStr,* and *CompareStringA* compared with other legitimate native modules, and it satisfies our assumption: a typical native module to extend Web server capability will not engage command execution and SQL queries into business data in a database.

## 6. Discussion

We can target to engage more different levels of detection of module backdoor in different web servers in the future. However, we need to seek more Malware samples for analysis to enhance the accuracy to detect benign IIS native modules and malicious backdoors. It is challenging because it is an emerging threat of APT, and most APT attack malware samples are confidential and private. We cannot easily accessible to them via public virus repositories.

The Module Backdoor with Web Server (IIS and Apache) is not as popular as ransomware spreads; it is always well-targeted against a particular organization. The entire campaign of attack is carried out with a thorough understanding of the IT technology, operation, and business nature of the target organization. The scanner we developed is just a beginning and the organization must set up a clear baseline that technology or query language must not be implemented casually for convenience according to the industry standard of web application design and deployment architecture. Otherwise, attackers can confuse the security administrator easily by deploying the module with “look-like” normal queries and operations in the backdoor.

With the scanner, an organization can add the keywords, and APIs should not be invoked and called normally in the Web Server module. It needs further efforts for this study of the API calls distribution in a host.

## 7. Conclusion

This research is to make organizations aware of cyber-security threats, which can be deployed in different layers inside a system architecture. We have found this emerging threat in IIS Native Module Backdoor as a crucial component in APT attack and our online business customer, which is an alarm we need to prepare to detect an insider threat and external attacker deploy it to the web server in different platforms and execute an unauthorized transaction and privileged operation. With the analysis of module backdoor installed in webserver, either on IIS or Apache, target organization can take TABMAX as a reference to detect this specific stealthy threat earlier.